# Fast GPU-Based Seismogram Simulation from Microseismic Events in Marine Environments Using Heterogeneous Velocity Models

Saptarshi Das, Xi Chen, and Michael P. Hobson

*Abstract*—A novel approach is presented for fast generation of synthetic seismograms due to microseismic events, using heterogeneous marine velocity models. The partial differential equations (PDEs) for the 3D elastic wave equation have been numerically solved using the Fourier domain pseudo-spectral method which is parallelizable on the graphics processing unit (GPU) cards, thus making it faster compared to traditional CPU based computing platforms. Due to computationally expensive forward simulation of large geological models, several combinations of individual synthetic seismic traces are used for specified microseismic event locations, in order to simulate the effect of realistic microseismic activity patterns in the subsurface. We here explore the patterns generated by few hundreds of microseismic events with different source mechanisms using various combinations, both in event amplitudes and origin times, using the simulated pressure and three component particle velocity fields *via* 1D, 2D and 3D seismic visualizations.

## I. Introduction

MONITORING of microseismic events is a growing area of research in computational geosciences that aims to determine spatial and temporal localization of small amplitude seismic events in the subsurface from the acoustic waves recorded by the geophones from the earth surface [1]. In a marine environment [2], the seismic traces recorded by the four component acoustic sensors (i.e. hydrophone measuring water pressure and tri-axial geophone measuring seabed vibration or three component particle velocity) are used to monitor the spontaneous seismic activity in the subsurface during and after hydrocarbon production and are often used for interpretation of the changing geological characteristics. In order to detect these sources reliably, the forward simulations, given an accurate heterogeneous velocity model of the geographic location, is a necessary step to understand the statistical characteristics of the wave-fields generated by the microseismic events. The forward simulation to generate synthetic seismic traces is quite challenging due to the heavy computational burden of solving elastic wave equation on a large number of grid points. Zhao *et al.* [3] have introduced a diffusive viscous wave equation with finite difference scheme on a 2D spatial model for simulation of synthetic seismograms. Sahimi and Allaei [4] introduced various heterogeneity correlation functions and compares the computational issues for solving the acoustic wave equation using finite difference, finite element and pseudo-spectral methods, highlighting the clear advantage of the latter. Synthetic seismogram generation for a 2D model has been shown in Phadke *et al.* [5], using an explicit finite difference predictor corrector scheme, parallelized with message passing interface (MPI) over multiple processors. Contrary to the real space methods, early developments in the Fourier domain solution of elastic wave equation can also be found in [6], [7] for the 2D and 3D models respectively.

The recent advent of GPU enabled numerical algorithms for solving PDEs on heterogeneous medium has revolutionized massive scalability of large 3D computational models that was infeasible, even few years ago. A detailed comparison of computational performance on different GPU hardware has been reported in Nickolls and Dally [8]. The GPU scalability of high level mathematical algorithms e.g. parallel Matlab codes as a popular and flexible choice of data analysis platform has been discussed in [9]–[12] along with various means to achieve it. Zhang *et al.* [13] compares the three most popular GPU computing tools in Matlab platform *viz.* parallel computing toolbox (PCT), Jacket and GPUmat, whereas performance comparison on various benchmark GPU computing problems have been reported in Shei *et al.* [14].

Wave propagation models have been implemented on GPUs with different discretization schemes e.g. finite difference [15][16][17][18], finite element [19][20][21][22], spectral element [23][24], discontinuous Galerkin method [25][26], and on multi-GPU clusters [27][28][29][30]. Computational efficiency of such models are compared on three different NVIDIA GPUs (C1060, C2050, M2090) in Zhou *et al.* [31] and in different size of models in Danek [32]. Traditionally for fast synthetic seismogram generation approximation methods like raytracing are widely used, but studying the source mechanism of the microseismic activity along with their position and origin time need accurate but computationally efficient forward simulation of the pressure and particle velocity fields, using the 3D heterogeneous models. Therefore solving full viscoelastic PDE is often a requirement compared

Manuscript received June 29, 2016; revised September 30, 2016; November 16, 2016; December 29, 2016 and January 5, 2017, accepted XX-XXXX. This work has been supported by the Shell Projects and Technology.

S. Das, X. Chen and M.P. Hobson are with the Cavendish Astrophysics Group, Department of Physics, University of Cambridge, Cambridge CB3 0HE, United Kingdom. (e-mail: {sd731, xc253, mph}@mrao.cam.ac.uk).

to the approximation methods, which leads to a massive computationally demanding step in terms of computing hardware, memory management, data processing and storage.

In this paper, starting from heterogeneous velocity models (specified by the density, primary (*P*) and secondary (*S*) soundwave velocities at each grid point), we generate a few hundreds of synthetic seismograms, as an effect of simulated microseismic activity at speculative locations in the subsurface. Then these seismograms are scaled by random amplitudes and translated for random origin times, followed by superposition of them, in order to simulate a realistic effect of microseismic activity, as observed in the receivers placed at the sea floor. Our numerical computation makes use of GPU enabled fast PDE solver for elastic wave equation, thus accelerating the simulations in a relatively cheaper way than the corresponding CPU versions.

The rest of the paper is organized as follows. Section II introduces the solution of elastic wave equations using heterogeneous velocity models. Section III reports the GPU computing performance for hundreds of sources and visualizes the synthetic seismograms as 1D, 2D, 3D images. Section IV shows combining these separately simulated seismograms with different amplitude, origin time, in order to generate realistic microseismic activity patterns. Events with different source mechanisms have been explored in section V. Cluster of events with various source mechanism have been shown in VI. The paper ends with the discussion, conclusion, and future scope of research in Section VII.

## II. Materials and Methods

### A. Solving the Elastic Wave Equation on GPUs for Fast Synthetic Seismic Trace Generation

We use here the elastic wave equation solver k-Wave [33], for the numerical solution using a given 3D heterogeneous model. The sensors and sources can be modelled with an arbitrary geometry and the simulations can be run faster on GPUs with single precision (32-bit representation) [34], [35]. The governing equations of the stress-strain relation for the propagation of compressional and shear (*P* and *S*) waves are given by the following equation (1) which is known as the Kelvin-Voigt model for viscoelastic materials

$$\sigma_{ij} = \lambda \delta_{ij} \varepsilon_{kk} + 2\mu \varepsilon_{ij} + \chi \delta_{ij} \frac{\partial \varepsilon_{kk}}{\partial x_j} + 2\eta \frac{\partial \varepsilon_{ij}}{\partial x_j}. \quad (1)$$

This extends the theory of Hooke's law and conservation of momentum for an elastic medium to the viscoelastic case with absorption. The stress/strain tensors $\{\sigma, \varepsilon\}$ are related to the particle velocity ($v_i$) and material density ($\rho$) by (2) (in Einstein Tensor notation) with $\delta_{ij}$ being the Kronecker delta.

$$\frac{\partial \sigma_{ij}}{\partial t} = \lambda \delta_{ij} \frac{\partial v_k}{\partial x_k} + \mu \left( \frac{\partial v_i}{\partial x_j} + \frac{\partial v_j}{\partial x_i} \right) + \chi \delta_{ij} \frac{\partial^2 v_k}{\partial x_k \partial t} + \eta \left( \frac{\partial^2 v_i}{\partial x_j \partial t} + \frac{\partial^2 v_j}{\partial x_i \partial t} \right),$$

$$\frac{\partial v_i}{\partial t} = \frac{1}{\rho} \frac{\partial \sigma_{ij}}{\partial x_j}. \quad (2)$$

The three component particle velocity ($v_i$) can also be recorded from the waves at specified sensor locations within the computational domain. The acoustic pressure (*p*) can then be derived from the stress tensor, using equation (3)

$$p = -\sigma_{ii}/3 = -(\sigma_{xx} + \sigma_{yy} + \sigma_{zz})/3. \quad (3)$$

Here, $\{\chi, \eta\}$ are the compressional and shear viscosity coefficients and $\{\lambda, \mu\}$ are the Lame parameters which are related to the compressional and shear sound speed of the material $\{c_p, c_s\}$ and density ($\rho$) through (4)

$$\mu = c_s^2 \rho, \quad (\lambda + 2\mu) = c_p^2 \rho. \quad (4)$$

Here, we have considered a lossless medium which can also be considered as frequency dependent absorption terms relating the sound velocity, density and the two viscosity terms $\{\chi, \eta\}$. In most of the geophysical simulation studies, the heterogeneous sound velocity and density model (together called the velocity model) are used for solving the forward problem of acoustic/elastic wave propagation through a viscoelastic medium [5], [3]. For the simulation of only the elastic case like in [6], [7], the viscous damping terms $\{\chi, \eta\}$ in (1)-(2) can be set to zero. The k-Wave solver reads the velocity model $\{c_p, c_s, \rho\}$ as a 3D array (specified in each grid point) of the computational domain and uses a Fourier domain pseudo-spectral method for solving spatial derivatives in the coupled PDEs in (2) and a leapfrog finite-difference scheme for time marching (with a spatial and temporal staggered grid arrangement) as detailed in Treeby *et al.* [33]–[35]. The pseudo-spectral method for computing spatial derivatives through fast Fourier transforms (FFTs) and its inverse (IFFT) makes the k-Wave solver computationally efficient and parallelizable over GPUs. The GPU parallelization of k-Wave solver transforms the 3D arrays in a custom datatype called *gpuArray* through Matlab's PCT, while performing the computation on single/double datatype objects in CUDA enabled NVIDIA GPUs. After the computation, the distributed arrays can be retrieved from multiple GPU cores using the *gather* function of PCT. Usually this method helps in getting sufficient scale up of the achievable computational performance using high level Matlab codes, containing several intricate mathematical functions, without retranslating the whole PDE solver as custom CUDA kernels [11]. The k-Wave solver also makes efficient use of the fast custom function *bsxfun* of PCT for elementwise binary operations for GPU computing, leveraging optimized libraries like FFT/IFFT [35] for spatial gradient calculation in the wavenumber domain. Although in certain benchmark cases Matlab simulation of elastic PDEs may be slower, however linear operations are sufficiently fast in Matlab's PCT on GPU nodes, because of the underlying high-performance libraries like BLAS, LAPACK, FFTW etc. [12] which the present method leverages on while also reducing code customization and development time. The k-Wave solver has been previously used in large (few km) scale 2D geophysical modelling for generating seismic data with visco-elastic wave propagation as reported in Guo *et al.* [36] where the difference with other similar seismic wave propagators have been shown to be negligibly small. This supports using k-Wave solver as a viable option for synthetic seismic data generation using GPUs.

Similar 3D PDE problems with traditional finite difference discretization schemes (even with an MPI parallelization over multiple cores) struggle for computational speed in a 3D heterogeneous forward model simulation using the elastic wave equation as in [5], [37]. The elastic wave equation based seismic trace generation using various combinations of moment tensors have been discussed in Li *et al.* [38], although the computational issues on a large number (generally few millions) of grid points have not been clearly addressed. In the present paper, we show a systematic methodology in which starting from fast forward model simulation with individual seismic sources (involving various elements of the stress tensor), at speculative locations of a heterogeneous marine geological model, one can simulate microseismic activity with multiple sources which is also scalable on larger geophysical models through state of the art GPU computing techniques.

Here the microseismic sources are modelled as delta functions at specified grid points. However, the simulation with discrete spatial delta functions may lead to oscillations in time domain due to the use of Fourier-domain pseudo-spectral method of calculating spatial derivatives as discussed in [35]. To avoid this, a smoothing operation on the initial pressure or stress distribution is carried out using a Blackman window in the spatial frequency domain in order to reduce higher spatial frequencies. This is equivalent to setting a smoothed delta function or input source wavelet in reflectivity function based seismogram calculation for layered earth model through the convolution operator [39]. However, smoothing function of the source in the k-Wave solver can be modified like Hanning, Blackman or some other windows as discussed in [35]. The effect of visco-elastic medium has also been reduced to an elastic case only, for the present geophysics modelling.

### B. Heterogeneous Velocity Models and Simulation Settings

Another important issue to worth considering is the numerical stability of the PDE solving scheme during the discretization which is usually given by the Courant-Fredrichs-Lewy (CFL) number, given by (5) in 1D.

$$CFL = c_{max} \Delta t / \Delta x \quad (5)$$

Apart from stability, the CFL number controls the minimum required temporal step-size to accurately integrate the PDE, given the spatial step-size and is therefore limited by the maximum sound speed $c_{max}$ along each direction. A small CFL (= 0.3 as suggested in [35]) would produce more accurate results at the cost of more computational burden and increased data storage. At the edges of the computational domains, 10 grid points were reserved to impose an absorbing boundary condition [40], through a perfectly matched layer (PML) to avoid reappearance of the waves leaving one side to the other due to the use of FFT in spatial (wavenumber) domain. Here we explore two 3D heterogeneous velocity models, both having the same size [41], as shown in Figure 1, the latter having more complex geometric structure, along the lateral direction. For both the models, the density ($\rho$) and P/S sound velocities ($c_p$, $c_s$) are specified in kg/m$^3$ and m/sec respectively, while strengths of the point sources (compressive or shear stress) are specified in Pascal. The measurements at the sensors as a solution of the PDE (2) are also in Pascal for the pressure field and m/sec for three particle velocity fields.

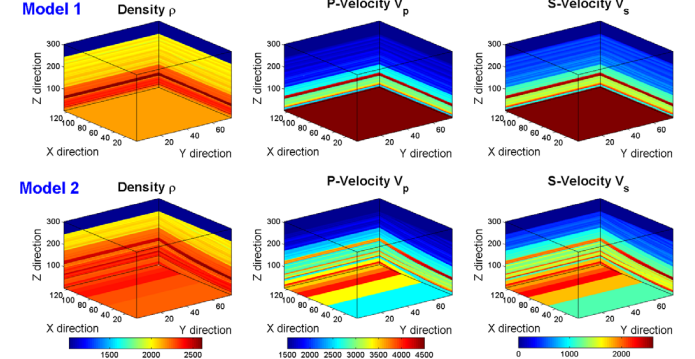

Figure 1: Two heterogeneous velocity models, comprising of the density ($\rho$), P-wave and S-wave velocity ($c_p$, $c_s$) at each grid point.

The simulations were run on a 3D domain of $N_x \times N_y \times N_z = 125 \times 75 \times 301 \approx 2.8219 \times 10^6$ grid points where the grid spacing in the three directions are given by $\Delta x = 12.5, \Delta y = 12.5, \Delta z = 10$ m. Therefore the physical dimension of the geological model is $1.550 \times 0.925 \times 3$ km$^3$. The idea here is to simulate synthetic seismic traces with sources lying at random positions in the rock volume, while the 4 component seismic sensors are placed at the sea floor. For both the velocity models, the location of the sensors are in every grid point ($N_x \times N_y = 125 \times 75 = 9375$ in total) at a fixed depth of $z = 244$. This is intuitively shown in Figure 2 by a scan through the two orthogonal central lines ($Nx/2$ and $Ny/2$), indicating that the sensors are located where the shear velocities just become zero. This signifies the interface between the rock volume and the sea bed in a marine environment. In both the velocity models, above the sensors, the remaining 56 grid points essentially model the effect of water column in the sea on top of the solid rock volume.

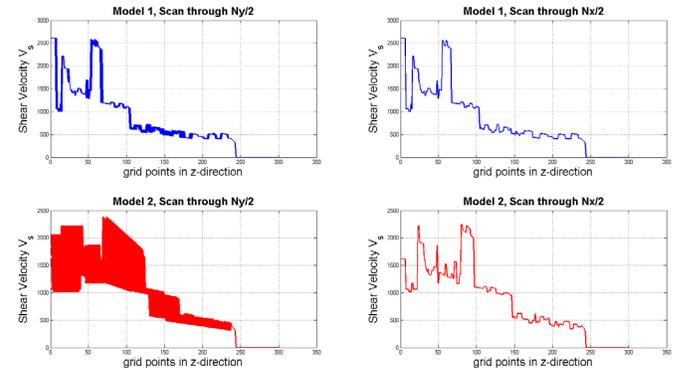

Figure 2: Sensor placement in the two velocity models at the grid point where the shear velocity just becomes zero ($z = 244$).

Following (5), given the spatial resolution and CFL = 0.3, the temporal resolution of the PDE is calculated as 0.5 msec. Also, the maximum and minimum time required by the sound wave to travel along $z$-direction from the farthest end of the model ($z = 1$) to the ocean bottom ($z = 244$) are calculated using the maximum and average sound velocity as 0.54 sec and 0.81 sec respectively. Therefore to ensure both numerical stability of the PDE solving scheme as well as capturing the information of the seismic waves due to a single explosive point source, even at the most distant location from the sensor, solving the elastic wave equation with a sampling time of $T_s$ = 0.0005 sec and a fixed time interval $T$ = 1 sec are sufficient.

## III. RESULTS OF SEISMIC TRACE SIMULATION ON GPUS

### A. Computational Performance on Two Different GPU Cards

Using the velocity model presented in the previous section with receivers placed at the sea floor, next we aim to activate pressure sources at speculative locations in the rock volume ($z$ < 244 for any $x$, $y$ location within the problem domain). We have used the Latin hypercube (LH) sampling to draw random source locations from the rock volume with unit strength (1 Pascal). The LH sampling explores maximum spatial volume covering maximum number of rows/columns in each dimension with a minimum number of samples and has been a popular choice in the field of geostatistics [42]. For both the heterogeneous velocity models, 2000 LH samples were drawn as possible source locations and also to compare the computational performances on two different GPU cards - Tesla C2075 and Tesla K20 as shown in Figure 3.

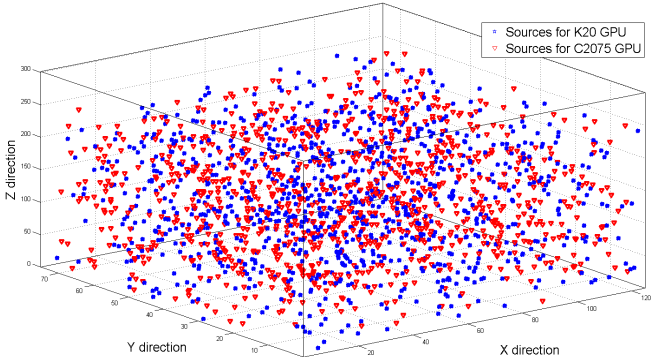

Figure 3: LH sample source locations in microseismic simulation on 2 GPUs.

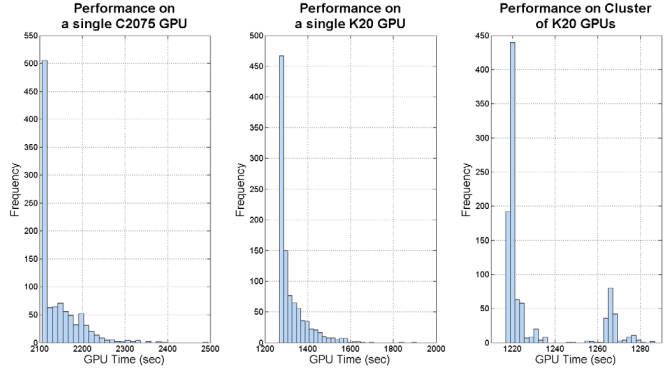

Figure 4: Distribution of computational time on two GPU cards – Tesla C2075, Tesla K20 and K20 GPU cluster with 10 simultaneous batches.

For the same forward simulation problem with a size of 2.8 million grid points, the CPU time (on a 64 bit Windows desktop PC with 16 GB memory and Intel I5, 3.3 GHz processor) for elastic wave equation solving takes 15 hours 41 mins for a single source position. Therefore for 2000 random source locations, the estimated CPU time would be 31367 hours ≈ 3.58 years which seems computationally infeasible. The same problem has been run on two different GPU cards *viz.* Tesla C2075 (with 5.5 GB memory, 1.1 GHz processor clock rate) and Tesla K20 (4.9 GB memory, 0.7 GHz processor clock rate). This makes the forward simulation ~27 times and ~43 times faster respectively as shown from the distribution of the computational time for the two GPU cards in Figure 4. Also there is data explosion in the forward seismic simulation process for integrating the PDE with a smaller time step. Each data file for single shot simulations, containing the acoustic pressure and three component particle velocity fields with a sampling time of $T_s$ = 0.5 msec in the 32-bit (single) precision occupies 262 MB memory per source on an average, totaling 524 GB for each of the velocity models.

For simulation using the velocity model 1, the computations were carried out in two GPU cards – K20 and C2075 in batches of 1000 sources, whereas for the velocity model 2, the Wilkes GPU cluster at the University of Cambridge has been used with batches assigned on 10 GPU cards simultaneously and the simulations have been parallelized for independent source locations with the same velocity model. The distribution of computational time for these three settings are compared in Figure 4. The single K20/C2075 GPU simulations were run in interactive mode of Matlab which is slightly slower, whereas the GPU cluster simulations were run in non-interactive batch mode which may have caused thinner runtime distribution than the former cases as in Figure 4.

The size of the velocity model has been chosen in such a way so that it matches a realistic computation time over a total simulation timeframe of $T$ = 1 sec, with a sampling time of $T_s$ = 0.5 msec, using both the Tesla K20 and Tesla C2075 GPUs. For simulation of larger velocity models, the problem domain can be divided in full depth ($z$ = 1 to 301 grid points) but reduced $x$ and $y$-direction length and solved separately, as the rock layer properties normally change more rapidly along the $z$-direction compared to the $x$ and $y$-direction. As revealed from Figure 4, the total time for 1000 samples of forward simulation is ~15 GPU days on Tesla K20 and ~25 GPU days on Tesla C2075, whereas on the K20 cluster with 10 parallel batches the total computation time is ~14 GPU days.

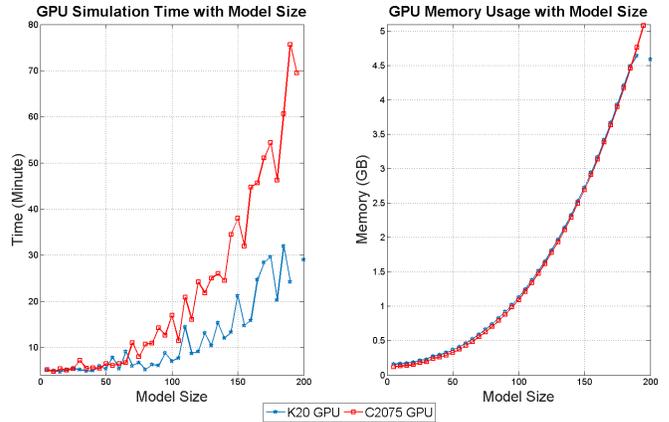

Figure 5: Scaling of simulation time and memory usage with increasing model size (grid point along a single axis) on two different GPUs. Model size $N$ represents a volume of $N^3$ grid points each with specified material properties.

Scaling of the forward simulation is also an important parameter on different GPUs and we here use a cubic model of size $N^3$ where $N_x = N_y = N_z = N$. Both the computational time and memory usage are shown in Figure 5, as a function of increasing size of the velocity model. Figure 5 also shows that the maximum size of the velocity models that fits well in a single GPU card has $N_x \times N_y \times N_z = 190^3 = 6.859 \times 10^6$ grid points. For larger velocity models, it overflows the GPU memory limiting the maximum size of a cube that can be accommodated for the forward simulation. However the real scale up between the K20 *vs.* C2075 GPU depends on the size and dimension of the velocity model, if it can be easily accommodated within a cube of ~$190^3$ grid points. Memory requirement on both the GPUs are found to be almost similar.

### B. Seismogram Wiggle Plots: 1D Visualization

The seismic traces for the pressure and particle velocity wave-fields are recorded in a fine temporal granularity as a solution of the PDE in (2). Now a single explosive source has been activated at the center of the volume (*Nx*/2, *Ny*/2, *Nz*/2). The elastic wave equation (2) updates the 6 stress tensor components and 3 component particle velocity at each time step. The propagation of normal and shear stress fields have been shown in the supplementary material for brevity.

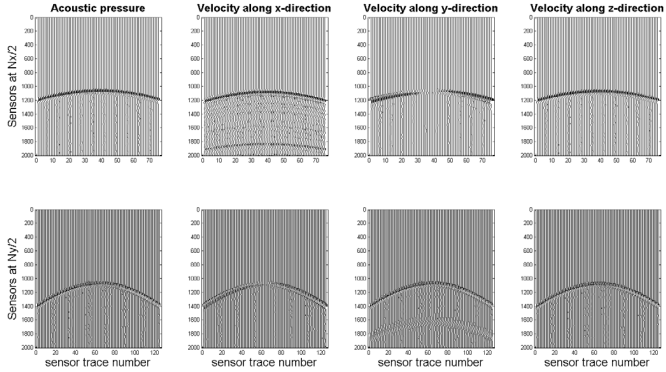

Figure 6: 1D depiction of seismogram trace wiggle plots for velocity model 1.

The resulting pressure and velocity fields after numerically solving the elastic wave equation on the heterogeneous medium with specified source strength and location can be visualized in various ways e.g. a popular option is 1D traces of pressure and particle displacement as wiggle plots, varying with the trace number as shown in Mulder [40]. Due to the 2D sensor placement at the sea floor, we have shown the 1D wiggle plots in Figure 6, for each trace along the central lines - *Nx*/2 and *Ny*/2 for the pressure and 3 component particle velocity fields for velocity model 1. Here the P/S-wave arrival times and seismogram patterns are expected to be different (predominantly found in the *x*-direction velocity) for the same source position with respect to that with the velocity model 2 and can be found in the supplementary material.

### C. Seismogram Plan Views: 2D Visualization

Since the sensors are placed in all the *x-y* grid points at a constant $z = 244$, each sensor essentially records a 4 component time series. Therefore at each time slice, the relative amplitudes amongst the sensors are a function of *x* and *y*-grid points ($s_{ij}(t), i \in [1, N_x], j \in [1, N_y]$) that can be represented in the form of time varying images of fixed size. For both the velocity models, two representative examples of the temporal evolution of pressure fields are shown in the supplementary material, over 1 sec of simulation time with an interval of 0.1 sec (staring from $T = 0.5$ sec).

### D. Seismogram Iso-surface Views: 3D Visualization

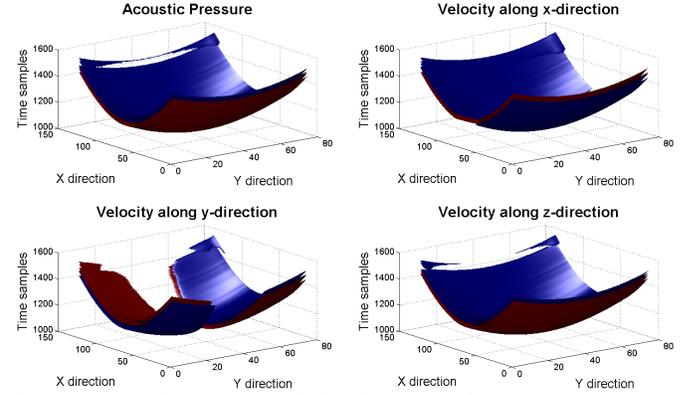

Figure 7: Iso-surface at an amplitude of ($\mu_s \pm 3\sigma_s$) for pressure and velocity fields for velocity model 2.

The four components measured by the sensors at different *x-y* location essentially capture the effect of propagating wave-field which has a geometric pattern for a single seismic source at a specified location. With a threshold on the wave-field amplitude on its 4 components using a ($\mu_s \pm 3\sigma_s$) criterion can uncover the strong positive/negative measurements in the form of a 3D surface, where $\{\mu_s, \sigma_s\}$ are the mean and standard deviation of the particular wave-field signal amplitudes amongst all the grid points. For model 2, the 3D iso-surface (at a constant amplitude) are shown in Figure 7 for pressure and three velocity fields. For a lesser heterogeneous velocity model like model 1, the generated patterns are expected to be smoother for a single microseismic source in the subsurface.

### IV. SIMULATION OF REALISTIC MICROSEISMIC EVENT PATTERNS BY COMBINING INDIVIDUAL SEISMIC TRACES

Results reported in the previous section explore the 1D, 2D and 3D imaging of the seismic traces due to a single point source at the center of the volume ($z = 150$) with unit strength (1 Pascal) and zero origin time ($T^{origin} = 0$). But in reality the strength, spatial location and origin time of the microseismic sources can take any arbitrary value, with multiple active sources having significant spatial and temporal proximity between them. In order to explore these complex situations, we superimpose the simulated wave-fields due to the sources at random locations as in Figure 3 in various different combinations. Since the elastic wave equation in (2) is linear in stress ($\sigma_{ij}$) and velocity ($v_i$) fields, the superposition principle holds for adding up the effect of individual sources with a scaled up/down strength. Due to the linear nature of the PDE, the individual wave-fields can also be time translated by a fixed amount to accommodate the effect of shifted origin times for different sources. In this section, we explore three different cases with increased modeling complexity, with different combinations of the number of sources ($N_e$), their positions, strengths and origin times, whereas the sensor locations and the velocity models are kept fixed.

## A. Case 1: Three Sources with Different Depth but Same Strength and Activated at the Same Origin Time

It is apparent that larger distances ($d$) between the sources and the receivers result the seismic traces fade away. In order to simulate such a realistic scenario, three unit sources (at $z$ = 160, 150, 80$^{th}$ grid point) have been activated at the same origin time $T_0$. The corresponding signal amplitudes recorded on the central receiver at ($Nx/2$, $Ny/2$) for these three events are shown in Figure 8 for both the velocity models. Attenuation of the maximum amplitude in the deeper sources are evident where the signal strength decreases with larger $d$.

The resulting wiggle plots for model 2 are shown in Figure 9. Due to different depth of these three sources, the arrival times of the wave-fronts are found distinctly different in the 1D seismic traces. The corresponding 2D plan views at different time slices are shown in the supplementary material for the pressure wave-field on both the velocity models.

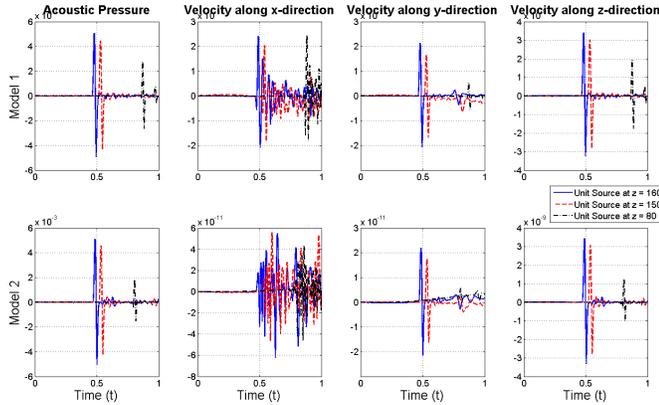

Figure 8: Amplitudes of seismic signals due to 3 unity strength events at ($Nx/2$, $Ny/2$) but different depth ($z$) activated at the same origin time.

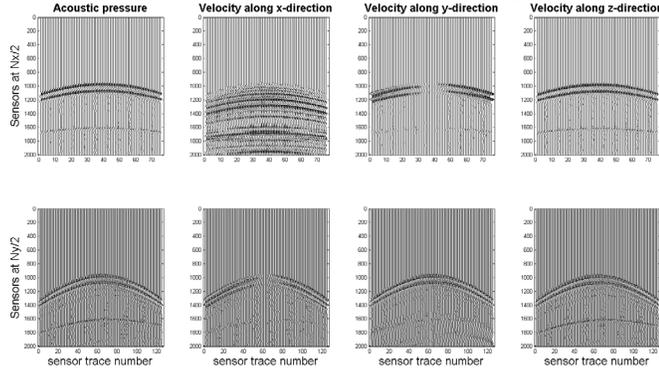

Figure 9: Seismogram wiggle plots for model 2 with 3 superimposed events at different depth but same strength and origin time.

## B. Case 2: Three Sources with Random Strength Activated at Random Origin Times

Compared to the example shown in case 1, here along with the random strength of the three sources, the origin times are also shifted by a random amount and then superimposed together. For implementing the random origin time, first the temporal length of the array is doubled and each seismic trace is shifted by a random amount (less than the signal length), followed by superimposition on the previous one. The rest of the entries after the shifted signal are zero padded. The 1D wiggle plots of the random origin times are shown in Figure 10 and the corresponding 2D plan views are shown in the supplementary material. This is an important case to explore as low amplitude sources near the receivers with recent origin time may be confusingly appear as stronger but deep sources with an older origin time.

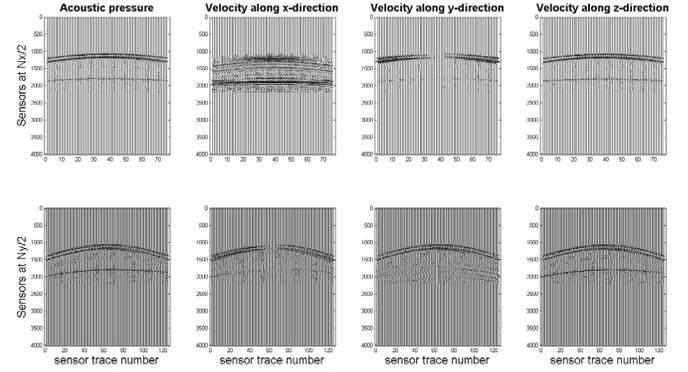

Figure 10: Seismogram wiggle plots for model 2 with 3 superimposed sources at different depth with random strength and random origin time.

## C. Case 3: Multiple Sources at Random Locations with Random Strength Activated at Random Origin Times

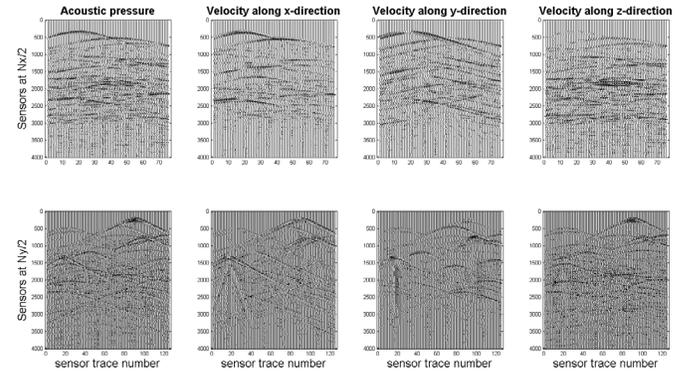

Figure 11: Seismogram wiggle plots for model 2 with 250 superimposed events with random amplitude and random origin time.

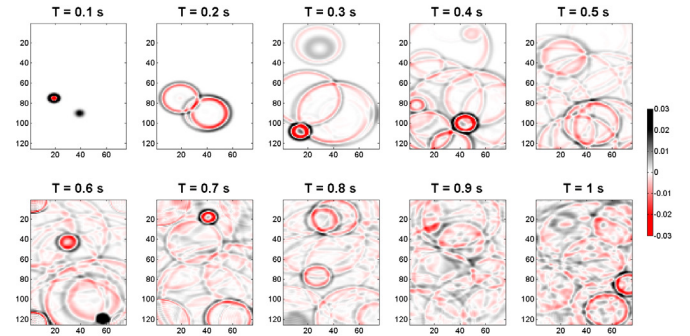

Figure 12: 2D plan view of the pressure field for model 2 with 250 sources with random amplitude and random origin time.

We now explore the effect of random source positions as shown in the LH samples in Figure 3 in addition to the previous cases. Along with random positions, the number of activated sources are also varied as $N_e$ = 10, 50, 250, 500, where both the strength and origin time for each sources have been randomly changed, as explored in the previous subsection. The seismic patterns generated by multiple superimposed sources ($N_e$ = 250) are shown in the 1D wiggle plots in Figure 11 and the corresponding 2D plan view in Figure 12. The rest of the seismic traces and the corresponding 2D plan views are shown in the supplementary material for brevity, with $N_e$ = 10, 50, 500 sources. It is apparent from Figure 11 and Figure 12 that the pressure fields corresponding to large number of sources, even identifying the number of sources become quite complicated. These recordings can often be confusingly appear as noisy seismograms, purely from a visual inspection of the superimposed traces with random event strengths, positions and origin times. However these complex patterns are generated from purely deterministic rules of superposition without any explicit noise term in the model. In the results in this section, we assume that all microseismic sources are explosive in nature without any shear component.

## V. EVENTS WITH DIFFERENT SOURCE MECHANISMS

### A. Basic Source Mechanisms for Microseismic Activity

Compared to the simulations reported in previous section, the actual microseismic sources may have complicated source mechanisms *viz*. with different shear ($\tau_{ij}$) and normal ($\sigma_{ii}$) stress components as characterized by the stress tensor (6).

$$\sigma = \begin{bmatrix} \sigma_{xx} & \tau_{xy} & \tau_{xz} \\ -\tau_{xy} & \sigma_{yy} & \tau_{yz} \\ -\tau_{xz} & -\tau_{yz} & \sigma_{zz} \end{bmatrix} \quad (6)$$

It is shown in [43], [38] that complex source mechanism can be mathematically decomposed in three basic type of sources *viz*. isotropic or explosive, double couple (DC) and compensated linear vector dipole (CLVD). The explosive source has already been discussed in the earlier sections where the diagonal components have the same sign (explosion or implosion). The other two basic source types, considered here can be represented as (7).

$$\sigma_{DC} = \begin{bmatrix} 0 & \tau_{xy} & 0 \\ -\tau_{xy} & 0 & 0 \\ 0 & 0 & 0 \end{bmatrix}, \sigma_{CLVD} = \begin{bmatrix} \sigma_{xx} & 0 & 0 \\ 0 & \sigma_{yy} & 0 \\ 0 & 0 & -2\sigma_{zz} \end{bmatrix} \quad (7)$$

### B. Wave-field Comparison in DC/CLVD vs. Explosive Source

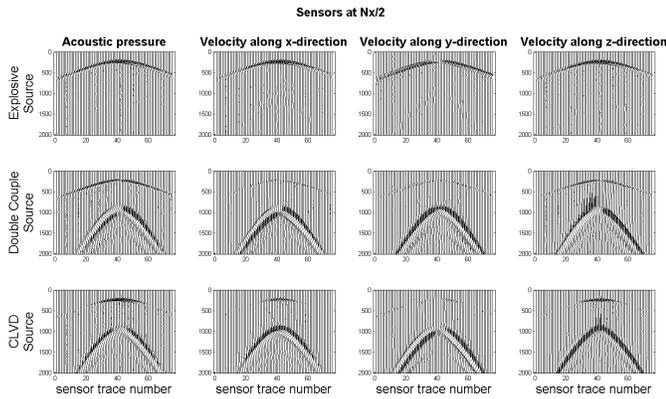

Figure 13: Seismogram wiggle plots for three type of source mechanism with the same source location ($x = 58$, $y = 41$, $z = 226$), recievers are alongs $Nx/2$. Stronger S-wave and weaker P-wave are observed in DC and CLVD sources.

Here the double couple source is considered to be in the *x-y* direction only, whereas in reality it can be along any other direction. The DC events model a shear source mechanism between two rock layers. The CLVD sources model a combination of tensile and compressive stress along different directions, often representing a sandwich model of two hard rock layers pressing a softer layer. It is shown in Li *et al.* [38] that DC and CLVD sources generate dominant S-waves whereas explosive sources have a dominant P-wave which is also evident from the wiggle plots in Figure 13. To compare the effect of these three source mechanisms on the generated wave-fields, a point source is activated at (58, 41, 226) grid point but with different components of the stress tensor using (7) to represent the explosive, DC and CLVD source mechanisms. The resulting 1D seismogram wiggles and 2D plan views for the acoustic pressure are shown in Figure 13 and Figure 14 respectively. The corresponding 4 component 3D iso-surface plots are shown in the supplementary material for brevity.

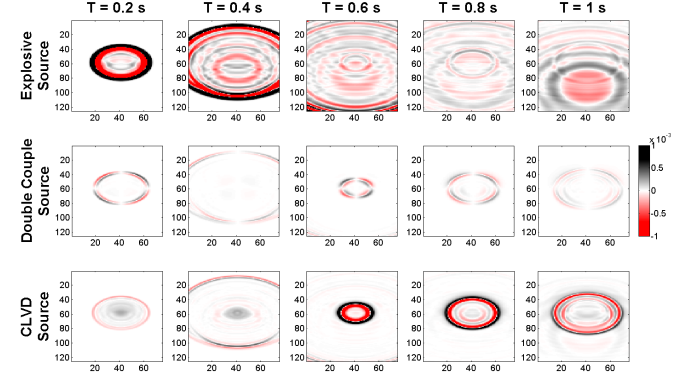

Figure 14: 2D plan view of the acoustic pressure for three types of source mechanisms with the same source location ($x = 58$, $y = 41$, $z = 226$). Note the different radiation patterns in explosive, DC and CLVD source mechanism.

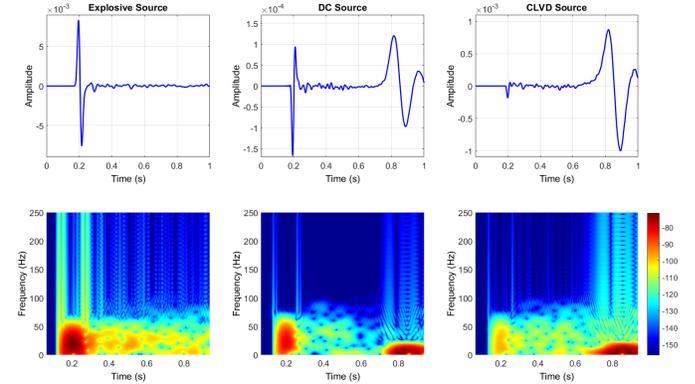

Figure 15: Time-frequency domain spectrogram representation of the seismic traces in the central receiver due to explosive, DC and CLVD sources.

In contemporary literature, microseismic activity has also been characterized using the time/frequency domain analysis for arrival time detection with tensile/shear slip [44], [45]. As quantitative measures, we here show the maximum frequency ranges in various cases of synthetic seismogram using explosive, DC and CLVD sources in Figure 15, as recorded in the central receiver. It is observed that although the frequency range of P-waves in explosive source may have components around 50 Hz and below but the dominant S-waves in DC/CLVD sources have a much lower spectrum ranging below 25 Hz which are in agreement with [44], [45]. The spectrograms were calculated using a $2^{10} = 1024$ point short time Fourier transform (STFT) with a sliding window of $2^6 = 256$ samples, as also applied in other studies related to the separation of seismic body waves [46].

## VI. SIMULATION OF CLUSTER OF MICROSEISMIC EVENTS AT RANDOM LOCATIONS WITH SPATIAL/TEMPORAL PROXIMITY

### A. Generating Cluster of Events in Space and Time

Many real microseismic studies reveal that the events are not really independent of each other (as assumed in the previous section), but there are significant spatial and temporal proximity/correlations amongst the sources within a cluster, generating a swarm like activity [47]. Also the events often occur at the interfaces between different rock layers where the material property changes quickly. Therefore out of the 1000 sampled locations, only a subset of events are chosen where the first/second spatial derivative of the density exceeds a certain threshold that can identify the edges between different rock layers. For both the velocity models, the 3D Laplacian of the density field (8) is calculated first as shown in Figure 16.

$$L = \nabla^2 \rho = \left(\partial^2 \rho / \partial x^2\right) + \left(\partial^2 \rho / \partial y^2\right) + \left(\partial^2 \rho / \partial z^2\right) \quad (8)$$

Next, only the sources above or below a fixed threshold of the Laplacian ($|L| > 0.1$) is retained as the active event locations, lying between different rock layers. In the velocity model 2, out of the first 500 LH samples in Figure 3, 82 locations match this criteria which are then represented as three (chosen *a priori*) non-overlapping groups or clusters [48], [49], using three different unsupervised learning algorithms.

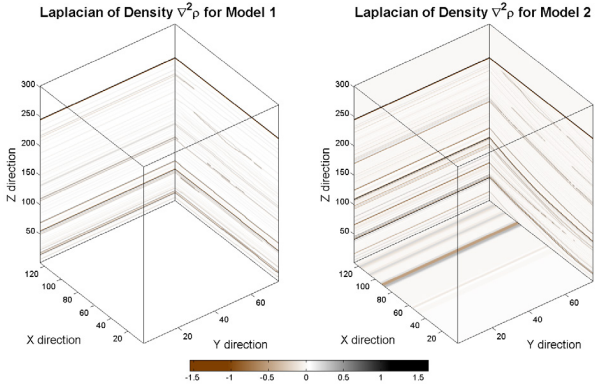

Figure 16: Laplacian $L = \nabla^2 \rho$ for detecting edges between different rock layers. Cracks and faults are likely to occur in these edges.

Firstly we apply the *k*-means clustering with 10 independent runs with random initial guess of the cluster centroids while minimizing the squared Euclidean distance performance measure. However the *k*-means algorithm prefers identifying spherical clusters, as it is based on only the mean value around each mode [50]. Another popular clustering technique known as the Gaussian Mixture Model (GMM) is employed next which is capable of finding ellipsoidal clusters as it adjusts the mean and covariance of each mode separately as well as the mixing proportions using the expectation maximization (EM) algorithm [50]. Similar to the *k*-means, 10 runs of the EM algorithm has been carried out on the same data with different start points in GMM to increase the robustness of finding similar clusters. As a third clustering algorithm for the sake of comparison, we use the hierarchical clustering which forms a tree like dendogram using the Euclidean distance criteria. The comparison of grouping the event locations in three clusters are shown in Figure 17 using these three clustering methods *viz. k*-means, GMM and hierarchical. It is observed that the *k*-means algorithm has been capable of grouping the event positions according to their depth, along which the variation in the rock properties are rapid, whereas the GMM algorithm although faithfully grouped cluster 1 and 3 but the variance of cluster 2 becomes large, resulting in spatial overlap of the clusters with the rest. The hierarchical clustering performs worst as it picks up isolated events at distant location as a separate clusters, as evident from the bottom panels of Figure 17. The purpose of the clustering methods here are to group the event locations in the forward simulation based on their spatial proximity, so that the origin times within a cluster can be made sufficiently close to simulate an effect of cascaded faults in the neighboring regions. It is also evident that the *k*-means algorithm assigns similar number of events in each cluster and groups the event locations with minimum degree spatial overlap, hence finally chosen over the other clustering methods for rest of the simulation study.

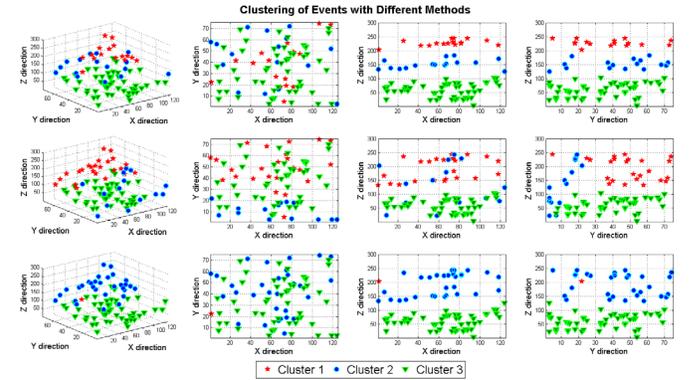

Figure 17: Source locations grouped in 3 clusters using various clustering algorithms (top) *k*-means, (middle) GMM, (bottom) hierarchical clustering.

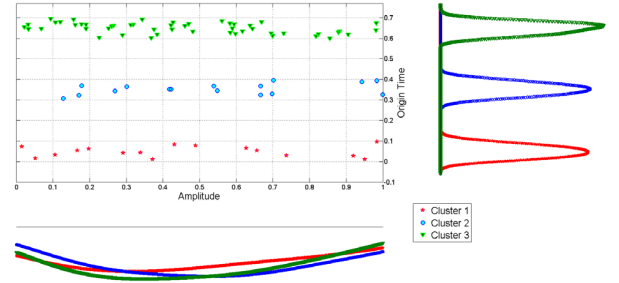

Figure 18: Scatter histogram of three cluster of events and distribution of their amplitudes and origin times. Origin times are separated by larger extent between clusters whereas event amplitudes are uniformly distributed.

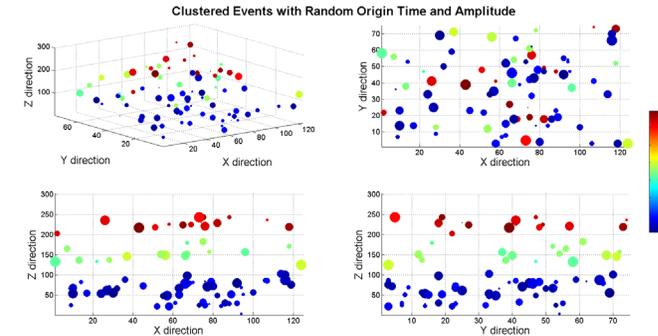

Figure 19: Event locations as a function of amplitude and origin time. Larger dots represent stronger amplitude and color represent shift in orgin time.

After relabeling the events in three clusters, their corresponding origin times ($T_i^{origin}$) are translated by 0.3 sec between these three clusters, while within the cluster they are translated by a smaller random amount with zero mean and 0.1 sec of standard deviation. The cluster of events nearest to the sea floor (cluster 1 in Figure 17) has been activated first, followed by the farthest sources in cluster 2 and cluster 3 as shown by their origin time *vs.* source amplitude scatter histograms in Figure 18. Here, the three clusters of origin times can be clearly identified with uniformly distributed strength of the events.

This particular case can be considered as a more realistic example where multiple events with spatial and temporal proximity are activated with smaller random origin time shift within a cluster but larger origin time shift between different clusters and a superimposed trace is recorded at the receivers. This helps physical modelling of cascaded growth of faults in the subsurface. These events are also plotted as a function of both their strength and the origin time in Figure 19 where the event sizes are proportional to their amplitudes and the colors represent differences in the respective origin times [51].

### B. Cluster of Events with Explosive Source Mechanism

Here we first assume that all the three cluster of events have explosive nature. For velocity model 2, the corresponding seimogram traces (as 1D wiggle plots) for the three cluster of events are shown in Figure 20 and the corresponding 2D plan views at different time slices in the top row of Figure 22.

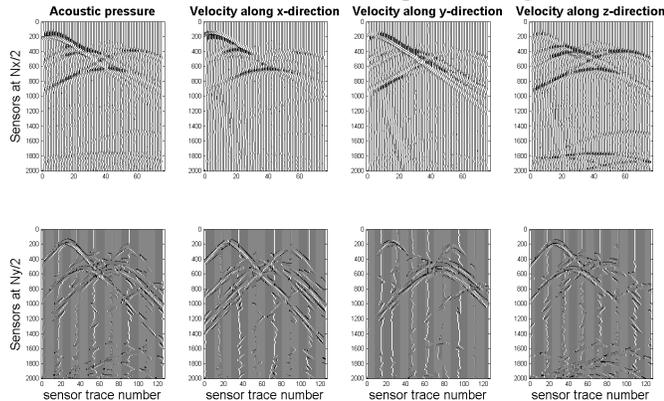
Figure 20: Seismogram wiggle plots for the four recorded components due to three clusters of explosive events in velocity model 2.

### C. Cluster of Events with Double Couple and CLVD Sources

Here we assume that all the three cluster of events have the same source mechanisms either double couple or CLVD type. Since the governing equations in (2) are linear in stress, the superposition of the seismic traces are not only valid for explosive sources but also for the shear sources where each point source may be characterized by a stress tensor in (6). To get a visual understanding of the generated wave-fields, the clustered event locations are considered fixed and with the same seed for random origin time and amplitude as reported in section VIA for different cases of microseismic events with the same source mechanisms in all the clusters.

The comparison of DC and CLVD sources in all the three clusters have been shown in the 1D seismogram wiggles for the pressure wave-field in Figure 21 with dominant S-waves.

The 2D plan views at different time slices for the DC/CLVD sources are compared with explosive ones in Figure 22. Here, the difference in polarity and radiation patterns can be identified clearly for the DC/CLVD events, as opposed to the explosive sources. In the generated wave-fields the appearance of positive and negative pressures often take the shape of concentric circles in the case of explosive and CLVD sources (with reverse polarity) but appears in opposite diagonals in the case of DC sources, as revealed from Figure 22.

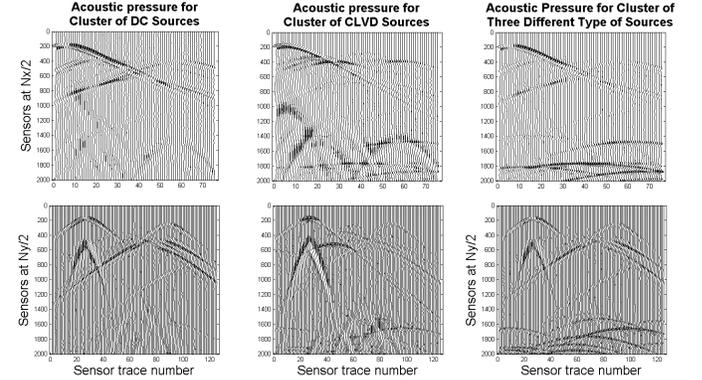
Figure 21: Comparison of seismogram wiggle plots due to three clusters of events with different source mechanism in velocity model 2. (left column) DC, (middle column) CLVD (right column) mixture of all the 3 sources types.

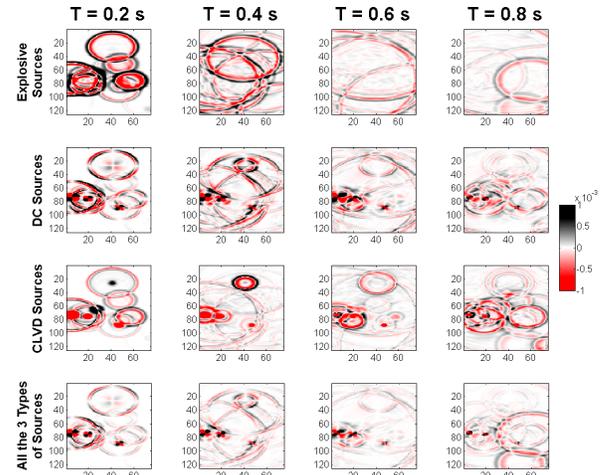
Figure 22: Comparison of 2D plan view of the pressure wavefield in velocity model 2, due to three cluster of events with same type (top three rows) and different type (bottom row) of microseismic sources.

### D. Different Source Mechanism in Different Clusters

We now explore the effect of having same source mechanism within the clusters but different mechanism in different clusters. Due to some degree of layered nature of the velocity models, the combined tensile and compressive stress sources are considered at the middle rock layers i.e. cluster 2 with CLVD sources. Due to the late arrival of S-waves which is predominantly found in the DC sources are considered in the layer close to the sea floor i.e. cluster 1 in Figure 17. The explosive sources generate dominant P-waves which travels much faster than the S-waves. Therefore cluster 3 or the deepest sources are considered to have explosive source mechanism. In this scenario, the waveforms from different type of sources are mixed together to generate highly complex wave-fields as shown in Figure 21 as 1D seismogram wiggles whereas the corresponding 2D plan views are shown in the bottom panel of Figure 22. The time-frequency domain spectrograms traced on the central receiver has been compared in Figure 23 for the clusters with the same type of sources, as shown in the previous subsections and also using combination of all the three types in this subsection. In the mixture of all the three type of source mechanisms, cluster 1 contains DC, cluster 2 contains CLVD, cluster 3 contains explosive sources. It is also evident from Figure 23 that the dominant power lies below the frequency range of $f_{max}$ = 50-80 Hz, depending on the different type of sources in the cluster of events.

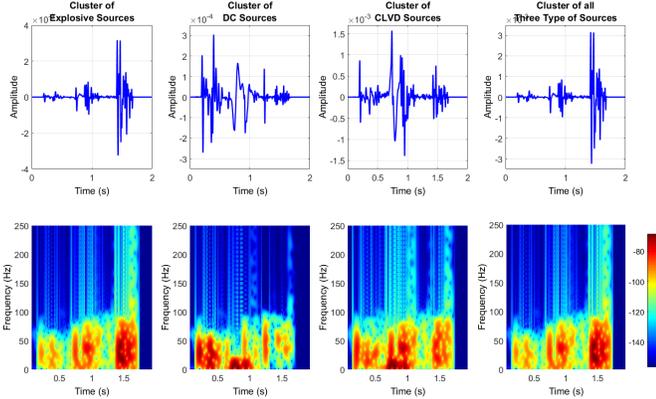

Figure 23: Time-frequency domain representation of the seismograms, traced in the central receiver due to cluster of explosive, DC, CLVD sources and mixture of all the three types of sources.

## VII. DISCUSSIONS AND CONCLUSION

### A. Achievements and Contributions over Existing Literature

A systematic methodology for fast forward simulation on GPUs are shown to generate synthetic seismic traces due to several cases of microseismic events in two realistic heterogeneous marine velocity models. Here we show that the use of GPU accelerated methods for computationally expensive forward seismic pattern generation due to microseismic events at random locations in the subsurface can be an efficient way, by using combination of few hundreds of separately simulated events. Although the seismic traces for an individual event looks simpler even for heterogeneous velocity models, in a realistic scenario there might be hundreds of simultaneous events taking place with random amplitudes and also shifted by random origin times. In such a scenario, the seismic traces may be quite complex, in terms of recorded pressure and three component particle velocity fields. This paper shows the following cases for the forward simulation of complex waveform modelling in microseismic response:

- Three different source mechanisms (explosive, double couple and compensated linear vector dipole events).
- Comparison of three different clustering methods ($k$-means, Gaussian mixture model and hierarchical clustering) to group the event locations with spatial proximity and then event origin time translation by smaller amount within cluster but by a larger amount between different clusters.
- Cluster of events with same and different type of source mechanism with random amplitude and origin time

Previous literatures (as introduced in section I) mostly attempted CPU/GPU based forward simulation of seismic wave propagation using single event on relatively simple velocity models and not necessarily always considering complex source mechanisms, which is an important factor in microseismic response modelling using elastic wave equation. Similar earlier works mainly focused on simplified 2D model with assumptions like layered earth, no lateral variation, special symmetry anisoropy, raytracing with only P/S-waves etc. Also reporting single shot simulation on GPU has been the most common e.g. [16], but carrying out bulk simulations using 1000s of sources and the resulting synthetic data storage, handling and memory management in itself is a challenging computational task which this paper deals with, as a contribution over existing literature.

Also, existing approaches mostly report 2D velocity models instead of the realistic 3D version e.g. in [16][17][24] and with only few layers in a 2D model as in [15][21] or often do not report on heterogeneous models at all e.g. [18][23][29][30], [31]. Danek [32] implemented a simplified acoustic wave propagation on GPU using only the pressure field without considering the full elastic wave equation and P/S-wave mode conversion. In Mu et al. [25], [26] a similar realistic 3D heterogeneous velocity model has been used with 0.03M and 0.3M cells respectively which is 22 times less than the computing requirements, needed for the present case having 6.86M cells, and even more for the simulation of few thousands of shots. Hence the reported simulations in this paper deals with sufficiently complex model along with larger computing requirements and associated complexities related to large volume of data storage and processing which has not been addressed in most of the previous literatures. Therefore the contribution of this paper is not only in the aspect of fast geophysical modelling using GPU cards but also processing of large volume of synthetic seismic data (>1 TB for both the models) to generate cluster of microseismic patterns with spatio-temporal proximity.

To the best of our knowledge, there is no literature that shows a step by step process of gradually building complexities in the microseismic forward modelling by combining several shots of single event simulation with various source mechanisms on a large heterogeneous velocity model, while also considering temporal and spatial proximity within the cluster of events with or without similar type of source mechanism. We also show that the complexities of the recorded wave-fields are generated using purely deterministic rules (scaling, translation and superposition of seismograms) by taking random samples from a heterogeneous velocity model, without considering a stochastic part in the forward modelling. Whereas in real seismic data low amplitude oscillations may often be interpreted as random background seismic noise which may be actually a result of weak deterministic events. Such a deterministic rule based microseismic response modelling opens up the scope of characterizing events when buried under significant amount of noise by separately considering the complexities due to the deterministic and stochastic parts of the forward modeling, within a Bayesian inference framework, in future.

### B. Need for a Fast Microseismic Response Simulation Method as the First Step for Detecting Events

In order to detect the microseismic events and quantify their characteristics reliably given the recorded 4 component seismic data, one viable option is to simulate template seismic response in a noiseless heterogeneous medium, and then applying machine learning techniques to detect them. Starting from unity amplitude single events at random locations, this paper shows various ways of complex seismic pattern generation with simple deterministic rules, e.g. effect of amplitude scaling, origin time shifting, superposition of multiple sources with scaling and shifting, grouping or clustering of correlated sources in space and time etc. which are quite similar to what is observed in a real hydrocarbon production field. Therefore, this paper can be considered as a first step towards the broader goal of reliable detection of microseismic events in the subsurface using different supervised learning algorithms. As a necessary first step, the seismic data generation process through numerical solution of the 3D elastic wave equation on few millions of grid points itself is quite computationally challenging task for generating such seismic event templates. The microseismic source parameters and corresponding noiseless synthetic seismic data can then be used to train an event detection algorithm for reliable prediction of unseen and often noisy data which may be pursued in a future research.

Therefore, with an aim of developing a Bayesian event detection framework for microseimic sources, a fast algorithm for approximate seismic template generation needs to be developed first, for faster computation of the likelihood. In a real field data, there might have many superimposed events, often added with complicated noise characteristics with joint spatio-temporal correlation structures. Separately using these superimposed seismograms as the deterministic part (often with a proposed event number and associated properties by a sampler) and a stochastic part (with known statistics of the background seismic noise) in the likelihood function may help in detecting many small amplitude superimposed events and often buried with correlated noise.

However, this paper do not focus on comparing numerical accuracy between various PDE solvers like in Phadke *et al.* [5], as there is no analytical solution possible for 3D elastic wave propagation through heterogeneous medium. As a result there will always be small numerical differences between different discretization schemes and specific implementations for the iterative PDE solvers as also shown in Guo *et al.* [36], including the k-Wave solver. However, these small inaccuracies in forward geophysics modelling with respect to the '*ideal noiseless physical response*' are also automatically considered in the likelihood function when matching with the real field data for microseismic event detection. Small increase in computational accuracy at the cost of higher computational burden and PDE discretization and numerical implementation specific complexities may not be worthwhile to explore here for a practical purpose, if approximate source modelling is the primary goal for event detection, using a Bayesian inference framework. This is even more justified in the presence of complicated noise characteristics in the receivers in a real field survey. Previously Poliannikov *et al.* [52]–[54] used a travel time based Bayesian inference, although the influence of background seismic noise on the travel time calculation may not have a simple linear and Gaussian nature. However this condition can be relaxed if the Bayesian inference can be framed rather in the real measurement i.e. the seismic domain where the noise can be assumed to have simpler statistical characteristics like known mean and covariance, across different receivers which is often considered using plethora of available covariance estimation techniques in geostatistics.

### C. Open Research Questions and Scope of Future Work

Future research may include the use of machine learning surrogate models to approximate the forward seismic simulations as a '*proxy*' for the GPU based simulations, and for fast likelihood calculation within a Bayesian microseismic event detection framework. Our method relies on the fact that large velocity models (including the extra PML grid points to impose absorbing boundary conditions) can be fitted well within the GPU memory which is a limitation to this study. In fact multi-GPU domain decomposition for larger 3D velocity models using distributed FFTs and multi-GPU communication aspects in a single shot simulations that does not fit in one GPU memory is an open research problem. One workable solution under the same framework could be decreasing the spatial resolution or upscaling of the model with a coarse grid (larger $\Delta x, \Delta y, \Delta z$) to reduce the number of grid points in the velocity model as adopted in [55], although as a consequence it may reduce the temporal resolution of the generated seismic data. Often finding a trade-off between the required temporal and spatial resolution may help reducing the number of grid points and hence the memory overflow problem, as well as the computing requirements, in a single shot simulation in GPUs.

Although the present simulation reports an ideal noiseless scenario, however in a real measured seismic data, distinguishing noise and low amplitude microseismic events is an open challenge. Ideally the number of occurring events are not known *a priori* and some assumption can be made for the event detection e.g. say there is known finite number of events ($N_e$) in the data, within a small time window or the event number can also be searched for, as an extra parameter in the inference process. Ideally, statistical characteristics of the background noise should be estimated from a long stream of data, so that even the presence of few low amplitude events do not significantly change the statistics of the noise. To scan for the microseismic events, a mixture of separately modelled simulated template microseismic response and laterally correlated seismic noise can be useful in the Bayesian likelihood calculation. However, noise estimation in real microseismic data in the context of event detection may be considered in a future study, as this paper mainly focuses on the fast GPU simulation of the forward problem, for synthetic noiseless seismic template generation, as required to calculate the likelihood function.

It is also important to note that the clustering is not used here to detect source positions or investigate the source mechanisms which is often known as the *inverse geophysical problem*. As discussed above that such an inverse problem to detect or quantify microseimic sources will typically need several thousands of forward simulations within a Bayesian sampling approach. The present paper focuses only on the first part of this broader objective to provide a fast solution for the forward geophysics modelling problem. The source mechanism can also be searched using a sampler or optimizer with a fixed or variable number of parameters in a future research. This may lead to a high dimensional sampling problem with unknown number of sources and associated properties like amplitude/stress components and origin times. Here, the clustering was carried out to show the generation of synthetic superimposed seismic waves with spatio-temporal proximity, as flexible templates in the forward model which can be used for comparison with real field data within the likelihood function.

This paper also focuses on the forward modelling of body seismic waves (P-wave and S-waves) only and does not include modelling of surface waves. Effect of different types of slow surface waves e.g. Rayleigh waves, Love waves and Stoneley waves can also be investigated in future.

APPENDIX

Additional high resolution images for the simulation results are provided in the supplementary material.

ACKNOWLEDGEMENT

This work used the Wilkes GPU cluster at the University of Cambridge High Performance Computing Service. GPU computing support from Greg Willatt and Dr. Stuart Rankin are gratefully acknowledged.

REFERENCES

[1] S. Mertl and E. Brückl, "Hazard Estimation of deep seated mass movements by microseismic monitoring," *ISDR20, Final Report*, 2008.
[2] S. S. Panahi, S. Ventosa, J. Cadena, A. Mànuel-Làzaro, A. Bermudez, V. Sallarès, and J. Piera, "A low-power datalogger based on compactflash memory for ocean bottom seismometers," *Instrumentation and Measurement, IEEE Transactions on*, vol. 57, no. 10, pp. 2297–2303, 2008.
[3] H. Zhao, J. Gao, and J. Zhao, "Modeling the Propagation of Diffusive-Viscous Waves Using Flux-Corrected Transport-Finite-Difference Method," *IEEE Journal of Selected Topics in Applied Earth Observations and Remote Sensing*, vol. 7, no. 3, pp. 838–844, 2014.
[4] M. Sahimi and S. M. V. Allaei, "Numerical simulation of wave propagation, part I: sequential computing," *Computing in Science & Engineering*, vol. 10, no. 3, pp. 66–75, 2008.
[5] S. Phadke, D. Bhardwaj, and S. Dey, "An explicit predictor-corrector solver with application to seismic wave modelling," *Computers & Geosciences*, vol. 26, no. 9, pp. 1053–1058, 2000.
[6] D. Kosloff, M. Reshef, and D. Loewenthal, "Elastic wave calculations by the Fourier method," *Bulletin of the Seismological Society of America*, vol. 74, no. 3, pp. 875–891, 1984.
[7] M. Reshef, D. Kosloff, M. Edwards, and C. Hsiung, "Three-dimensional elastic modeling by the Fourier method," *Geophysics*, vol. 53, no. 9, pp. 1184–1193, 1988.
[8] J. Nickolls and W. J. Dally, "The GPU computing era," *Micro, IEEE*, vol. 30, no. 2, pp. 56–69, 2010.
[9] S. Samsi, V. Gadepally, and A. Krishnamurthy, "MATLAB for signal processing on multiprocessors and multicores," *Signal Processing Magazine, IEEE*, vol. 27, no. 2, pp. 40–49, 2010.
[10] J. W. Suh and Y. Kim, *Accelerating MATLAB with GPU computing: A primer with examples*. Elsevier, 2014.
[11] Y. Altman, *Accelerating MATLAB performance: 1001 tips to speed up MATLAB programs*. CRC Press, 2015.
[12] M. Szymczyk and P. Szymczyk, "Matlab and Parallel Computing," *Image Processing & Communications*, vol. 17, no. 4, pp. 207–216, 2012.
[13] B. Zhang, S. Xu, F. Zhang, Y. Bi, and L. Huang, "Accelerating matlab code using gpu: A review of tools and strategies," in *Artificial Intelligence, Management Science and Electronic Commerce (AIMSEC), 2011 2nd International Conference on*, 2011, pp. 1875–1878.
[14] C.-Y. Shei, P. Ratnalikar, and A. Chauhan, "Automating GPU computing in MATLAB," in *Proceedings of the international conference on Supercomputing*, 2011, pp. 245–254.
[15] D. Michéa and D. Komatitsch, "Accelerating a three-dimensional finite-difference wave propagation code using GPU graphics cards," *Geophysical Journal International*, vol. 182, no. 1, pp. 389–402, 2010.
[16] F. Rubio, M. Hanzich, A. Farrés, J. De La Puente, and J. M. Cela, "Finite-difference staggered grids in GPUs for anisotropic elastic wave propagation simulation," *Computers & Geosciences*, vol. 70, pp. 181–189, 2014.
[17] J. Francés, S. Bleda, A. Márquez, C. Neipp, S. Gallego, B. Otero, and A. Beléndez, "Performance analysis of SSE and AVX instructions in multi-core CPUs and GPU computing on FDTD scheme for solid and fluid vibration problems," *The Journal of Supercomputing*, vol. 70, no. 2, pp. 514–526, 2014.
[18] D. Unat, J. Zhou, Y. Cui, S. B. Baden, and X. Cai, "Accelerating a 3D Finite-Difference Earthquake Simulation with a C-to-CUDA Translator," *Computing in Science & Engineering*, vol. 14, no. 3, pp. 48–59, 2012.
[19] D. Komatitsch, G. Erlebacher, D. Göddeke, and D. Michéa, "High-order finite-element seismic wave propagation modeling with MPI on a large GPU cluster," *Journal of Computational Physics*, vol. 229, no. 20, pp. 7692–7714, 2010.
[20] D. Komatitsch, D. Michéa, and G. Erlebacher, "Porting a high-order finite-element earthquake modeling application to NVIDIA graphics cards using CUDA," *Journal of Parallel and Distributed Computing*, vol. 69, no. 5, pp. 451–460, 2009.
[21] P. Huthwaite, "Accelerated finite element elastodynamic simulations using the GPU," *Journal of Computational Physics*, vol. 257, pp. 687–707, 2014.
[22] M. Rietmann, P. Messmer, T. Nissen-Meyer, D. Peter, P. Basini, D. Komatitsch, O. Schenk, J. Tromp, L. Boschi, and D. Giardini, "Forward and adjoint simulations of seismic wave propagation on emerging large-scale GPU architectures," in *Proceedings of the International Conference on High Performance Computing, Networking, Storage and Analysis*, 2012, p. 38.
[23] D. Komatitsch, D. Göddeke, G. Erlebacher, and D. Michéa, "Modeling the propagation of elastic waves using spectral elements on a cluster of 192 GPUs," *Computer Science-Research and Development*, vol. 25, no. 1–2, pp. 75–82, 2010.
[24] S. D. Walsh, M. O. Saar, P. Bailey, and D. J. Lilja, "Accelerating geoscience and engineering system simulations on graphics hardware," *Computers & Geosciences*, vol. 35, no. 12, pp. 2353–2364, 2009.
[25] D. Mu, P. Chen, and L. Wang, "Accelerating the discontinuous Galerkin method for seismic wave propagation simulations using the graphic processing unit (GPU)—single-GPU implementation," *Computers & Geosciences*, vol. 51, pp. 282–292, 2013.
[26] D. Mu, P. Chen, and L. Wang, "Accelerating the discontinuous Galerkin method for seismic wave propagation simulations using multiple GPUs with CUDA and MPI," *Earthquake Science*, vol. 26, no. 6, pp. 377–393, 2013.
[27] D. Komatitsch, "Fluid-solid coupling on a cluster of GPU graphics cards for seismic wave propagation," *Comptes Rendus Mécanique*, vol. 339, no. 2, pp. 125–135, 2011.
[28] R. M. Weiss and J. Shragge, "Solving 3D anisotropic elastic wave equations on parallel GPU devices," *Geophysics*, vol. 78, no. 2, pp. F7–F15, 2013.
[29] T. Okamoto, H. Takenaka, T. Nakamura, and T. Aoki, "Accelerating large-scale simulation of seismic wave propagation by multi-GPUs and three-dimensional domain decomposition," *Earth, Planets and Space*, vol. 62, no. 12, pp. 939–942, 2010.
[30] Y. Zhou, S. Song, T. Dong, and D. A. Yuen, "Seismic wave propagation simulation using accelerated support operator rupture dynamics on multi-gpu," in *Computational Science and Engineering*

[30] (CSE), 2011 IEEE 14th International Conference on, 2011, pp. 567–572.
[31] J. Zhou, D. Unat, D. J. Choi, C. C. Guest, and Y. Cui, "Hands-on performance tuning of 3D finite difference earthquake simulation on GPU fermi chipset," *Procedia Computer Science*, vol. 9, pp. 976–985, 2012.
[32] T. Danek, "Seismic wave field modeling with graphics processing units," in *Computational Science-ICCS 2009*, Springer, 2009, pp. 435–442.
[33] B. E. Treeby, J. Jaros, D. Rohrbach, and B. Cox, "Modelling elastic wave propagation using the k-wave matlab toolbox," in *Ultrasonics Symposium (IUS), 2014 IEEE International*, 2014, pp. 146–149.
[34] B. E. Treeby and B. T. Cox, "k-Wave: MATLAB toolbox for the simulation and reconstruction of photoacoustic wave fields," *Journal of Biomedical Optics*, vol. 15, no. 2, pp. 021314–021314, 2010.
[35] B. Treeby, B. Cox, and J. Jaros, "k-Wave A MATLAB toolbox for the time domain simulation of acoustic wave fields User Manual," 2012.
[36] P. Guo, G. A. McMechan, and H. Guan, "Comparison of two viscoacoustic propagators for Q-compensated reverse time migration," *Geophysics*, vol. 81, no. 5, pp. S281–S297, 2016.
[37] H. Igel, P. Mora, and B. Riollet, "Anisotropic wave propagation through finite-difference grids," *Geophysics*, vol. 60, no. 4, pp. 1203–1216, 1995.
[38] H. Li, R. Wang, and S. Cao, "Microseismic forward modeling based on different focal mechanisms used by the seismic moment tensor and elastic wave equation," *Journal of Geophysics and Engineering*, vol. 12, no. 2, p. 155, 2015.
[39] W. A. Mousa and A. A. Al-Shuhail, *Processing of Seismic Reflection Data Using MATLAB$^{TM}$*, vol. 5, no. 1. Morgan & Claypool Publishers, 2011, pp. 1–97.
[40] W. Mulder, "Experiments with Higdon's absorbing boundary conditions for a number of wave equations," *Computational Geosciences*, vol. 1, no. 1, pp. 85–108, 1997.
[41] G. S. Martin, R. Wiley, and K. J. Marfurt, "Marmousi2: An elastic upgrade for Marmousi," *The Leading Edge*, vol. 25, no. 2, pp. 156–166, 2006.
[42] M. Stein, "Large sample properties of simulations using Latin hypercube sampling," *Technometrics*, vol. 29, no. 2, pp. 143–151, 1987.
[43] V. Vavrycuk, "Moment tensor decompositions revisited," *Journal of Seismology*, vol. 19, no. 1, pp. 231–252, 2015.
[44] C. Levy, D. Jongmans, and L. Baillet, "Analysis of seismic signals recorded on a prone-to-fall rock column (Vercors massif, French Alps)," *Geophysical Journal International*, vol. 186, no. 1, pp. 296–310, 2011.
[45] D. W. Eaton, M. van der Baan, B. Birkelo, and J.-B. Tary, "Scaling relations and spectral characteristics of tensile microseisms: Evidence for opening/closing cracks during hydraulic fracturing," *Geophysical Journal International*, p. ggt498, 2014.
[46] R. H. Herrera, J. B. Tary, M. van der Baan, and D. W. Eaton, "Body wave separation in the time-frequency domain," *IEEE Geoscience and Remote Sensing Letters*, vol. 12, no. 2, pp. 364–368, 2015.
[47] D. Becker, T. Meier, M. Rische, M. Bohnhoff, and H.-P. Harjes, "Spatio-temporal microseismicity clustering in the Cretan region," *Tectonophysics*, vol. 423, no. 1, pp. 3–16, 2006.
[48] G. Adelfio, M. Chiodi, A. D'Alessandro, D. Luzio, G. D'Anna, and G. Mangano, "Simultaneous seismic wave clustering and registration," *Computers & Geosciences*, vol. 44, pp. 60–69, 2012.
[49] D. Fagan, K. van Wijk, and J. Rutledge, "Clustering revisited: A spectral analysis of microseismic events," *Geophysics*, vol. 78, no. 2, pp. KS41–KS49, 2013.
[50] S. Rogers and M. Girolami, *A first course in machine learning*. Chapman & Hall, 2011.
[51] L. Eisner, S. Williams-Stroud, A. Hill, P. Duncan, and M. Thornton, "Beyond the dots in the box: Microseismicity-constrained fracture models for reservoir simulation," *The Leading Edge*, vol. 29, no. 3, pp. 326–333, 2010.
[52] O. V. Poliannikov, M. Prange, A. E. Malcolm, and H. Djikpesse, "Joint location of microseismic events in the presence of velocity uncertainty," *Geophysics*, vol. 79, no. 6, pp. KS51–KS60, 2014.
[53] O. V. Poliannikov, M. Prange, A. Malcolm, and H. Djikpesse, "A unified Bayesian framework for relative microseismic location," *Geophysical Journal International*, p. ggt119, 2013.
[54] O. V. Poliannikov, M. Prange, H. Djikpesse, A. E. Malcolm, and M. Fehler, "Bayesian inversion of pressure diffusivity from microseismicity," *Geophysics*, vol. 80, no. 4, pp. M43–M52, 2015.
[55] N. Gold, S. Shapiro, S. Bojinski, and T. Müller, "An approach to upscaling for seismic waves in statistically isotropic heterogeneous elastic media," *Geophysics*, vol. 65, no. 6, pp. 1837–1850, 2000.